\documentclass[a4paper]{article}

\usepackage{INTERSPEECH2019}
\usepackage[english]{babel}
\title{Automatic Spelling Correction with Transformer for CTC-based End-to-End Speech Recognition}

\name{Shiliang Zhang, Ming Lei, Zhijie Yan}
\address{Machine Intelligence Technology, Alibaba Group}
\email{\{sly.zsl, lm86501, zhijie.yzj\}@alibaba-inc.com}

\begin{document}
	
\maketitle
\begin{abstract}
Connectionist Temporal Classification (CTC) based end-to-end speech recognition system usually need to incorporate an external language model by using WFST-based decoding in order to achieve promising results. This is more essential to Mandarin speech recognition since it owns a special phenomenon, namely \emph{homophone}, which causes a lot of substitution errors. The linguistic information introduced by language model will help to distinguish these substitution errors. In this work, we propose a transformer based spelling correction model to automatically correct errors especially the substitution errors made by CTC-based Mandarin speech recognition system. Specifically, we investigate using the recognition results generated by CTC-based systems as input and the ground-truth transcriptions as output to train a transformer with encoder-decoder architecture, which is much similar to machine translation. Results in a 20,000 hours Mandarin speech recognition task show that the proposed spelling correction model can achieve a CER of 3.41\%, which results in 22.9\% and 53.2\% relative improvement compared to the baseline CTC-based systems decoded with and without language model respectively.

\end{abstract}
\noindent\textbf{Index Terms}: speech recognition, spelling correction, CTC, End-to-End, Transformer

\section{Introduction}
Conventional hybrid DNN-HMM based speech recognition system usually consists of acoustic, pronunciation and language models. These components are trained separately, each with a different objective, and then combined together during model inference. Recent works in this area attempt to rectify this disjoint training problem and simplify the training process by building speech recognition system in the so-called end-to-end framework \cite{graves2014towards,chorowski2014end,hannun2014deep,chorowski2015attention,chan2016listen,amodei2016deep,bahdanau2016end,kim2017joint,chiu2018state}. Two popular approaches for this are the Connectionist Temporal Classification (CTC) \cite{graves2006connectionist} and attention-based encoder-decoder models \cite{bahdanau2014neural}. Both methods regard speech recognition as a sequence-to-sequence mapping problem and address the problem of variable-length input and output sequences. CTC uses intermediate label representation allowing repetitions of labels and occurrences of blank label to identify less informative frames, which enables CTC-based acoustic models to automatically learn the alignments between speech frames and target labels. On the other hand, attention based models use an attention mechanism to perform alignment between acoustic frames and recognized symbols. Both methods do not require frame-level training targets, which simplifies the training process of speech recognition system.

CTC assumes that the label outputs are conditionally independent of each other, which can be seen as an acoustic-only model. Although CTC-based models can directly generate the recognition results by using the greedy search decoding \cite{graves2006connectionist}, it's better to incorporate an external language model at the character or word level by using the WFST-based decoding \cite{miao2015eesen,li2015towards}. On the other hand, attention-based models with \emph{encoder-decoder} framework can jointly learn acoustic, pronunciation and language models. As a result, it is widely observed that attention-based models will achieve better performance than CTC-based models when decoded without external language model \cite{bahdanau2016end}. However, the language model component in attention-based models is only trained on transcribed audio-text pairs. Further improvements can achieve by incorporating an external language model at inference time \cite{kannan2018analysis}.

Mandarin is a tonal language with a special phenomenon, namely \emph{homophone}, which many Chinese characters share the same pronunciation. As a result, the substitution errors are the dominant error made by Mandarin speech recognition system \cite{zhang2019deep}. These substitution errors require linguistic information to distinguish effectively. Thereby, the language model is essential to CTC-based models for Mandarin speech recognition.  As shown in \cite{li2015towards}, the performance gap between CTC-based models decoded with and without external language model is huge. However, the language model accompanied with CTC-based acoustic models is usually the n-gram language model, which has limited history context information. Further improvement can be obtained by using N-best rescoring with an RNN-LM \cite{mikolov2010recurrent,liu2016two}.

In this work, we propose a transformer \cite{vaswani2017attention} based spelling correction model to automatically correct some errors made by the CTC-based speech recognition system. Specifically, we investigate using the recognition results generated by CTC-based systems as input and the ground-truth transcriptions as output
to train a transformer with encoder-decoder architecture, which is similar to machine translation. During inference, the spelling correction model takes the preliminary
recognition results as input and generates the final results with greedy search. We have investigated various CTC-based systems as front-end: different acoustic modeling units (syllable, character-2k, character-4k, character-6k), different optimization criteria (CTC, CTC-sMBR) and decoding methods (greedy search, WFST search). Moreover, we have proposed to extend the diversity of training data by using N-best lists and used the SGDR \cite{loshchilov2016sgdr} optimization, which will significantly improve the performance. 

We have evaluated our proposed approach on a 20,000 hours Mandarin speech recognition task that consists of about 20 millions paired sentences. Our results show that the proposed spelling correction models can improve the performance of CTC system with greedy search from 7.28\% to 4.89\% in character error rate (CER). And it can further improve to 4.21\% by extending N-best lists as training data. As a comparison, the performance of a well-trained CTC-sMBR system using WFST-based decoding with external word-level language model is 4.42\%. Moreover, by jointly using character based acoustic modeling units, DFSMN-CTC-sMBR acoustic model, WFST-based decoding and N-best data expansion, the proposed spelling correction models can achieve a CER of 3.41\%, which results in a 22.9\% relative improvement. Our analysis show that the transformer based spelling correction model can significantly reduce substitution errors in recognition results, due to it can utilize the sentence-level linguistic information.

\section{Related Works}
Automatic correction of recognition errors is crucial not only to improve the performance of ASR system but also to avoid the propagation of errors to the post
process (e.g. machine translation, natural language processing). \cite{errattahi2018automatic} presents an overview of previous work on error correction for ASR.
However, most of researches were limited to the detection \cite{zhou2005data,allauzen2007error,pellegrini2009error} and just few researches addressed the correction process of ASR errors. In \cite{sarma2004context}, it built an ASR errors detector and corrector using co-occurrence analysis. \cite{bassil2012asr} proposed a post-editing ASR errors correction method based on Microsoft N-gram dataset. More recently, work in \cite{zenkel2017comparison,guo2019spelling} proposed to use the attention based sequence-to-sequence model to automatically correct the ASR errors, which is much similar to our work.

\section{Our Approach}
\begin{figure}[t]
	\centering
	\includegraphics[width=1.0\linewidth]{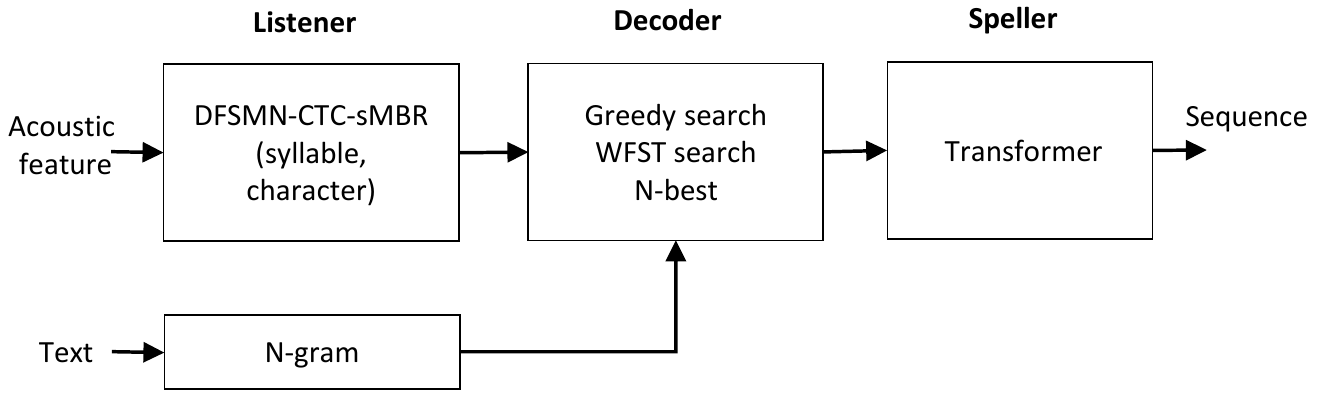}
	\caption{Overall system of the proposed approach.}
	\label{fig:system}
\end{figure}
Figure \ref{fig:system} demonstrates the overall system of the proposed approach, which consists of three components: \emph{listener}, \emph{decoder} and \emph{speller}. For listener, we use the DFSMN-CTC-sMBR  \cite{zhang2019deep} based acoustic model. As to decoder, we compare the greedy search \cite{graves2006connectionist} and WFST search \cite{miao2015eesen} based decoding strategies to generate preliminary recognition results given static sequence of probabilities generated by the listener. Moreover, we also investigate how to extend the diversity of preliminary recognition results with candidate N-best. Finally, the outputs generated by the decoder are used to train a transformer based speller. 

\subsection{Listener}
\subsubsection{DFSMN-CTC-sMBR}
\begin{figure}[t]
	\centering
	\includegraphics[width=0.8\linewidth]{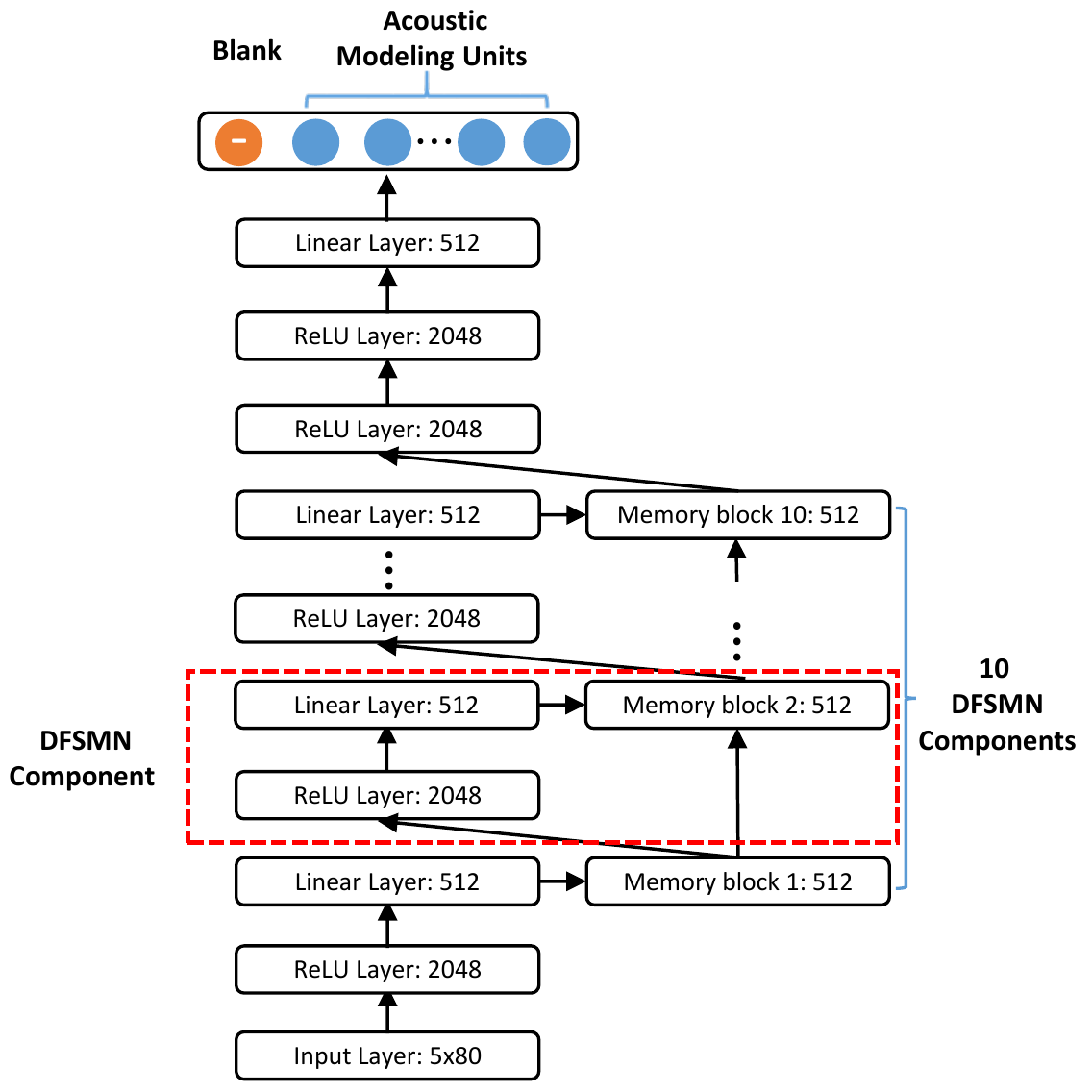}
	\caption{Illustration of DFSMN-CTC-sMBR acoustic model.}
	\label{fig:dfsmn}
\end{figure}
DFSMN \cite{zhang2018deep} is an improved FSMN \cite{zhang2015feedforward} structure that enables to build extreme deep architecture by introducing skip connections. As shown in Figure \ref{fig:dfsmn}, it is a DFSMN with 10 DFSMN-components followed by 2 fully-connected ReLU layers, a linear layer and a softmax output layer. For CTC based acoustic model, the softmax output corresponds to the acoustic modeling units and a \emph{blank} unit. The key element in DFSMN is the learnable FIR-like memory blocks, which are used to encode long context information into fixed-size representation. As a result, DFSMN is able to model the long-term dependency in sequential signals while without using recurrent feedback. The operation in $\ell$-th memory block takes the following form:
\begin{equation}\label{eq.DFSMN}
{\bf m}_t^{\ell}={\bf m}_t^{\ell  - 1}+ {\bf p}^\ell_t+\sum\limits_{i = 0}^{N^\ell_1} {\bf a}_i^{\ell}  \odot {\bf p}^{\ell}_{t - s_1*i} + \sum\limits_{j = 1}^{N^\ell_2} {\bf c}_j^{\ell} \odot {\bf p}^{\ell}_{t + s_2*j}
\end{equation}
Here, ${\bf p}_t^\ell$ denote the outputs of the linear projection layer and ${\bf m}_t^{\ell}$  denotes the output of the memory block. $N^\ell_1$ and $N^\ell_2$ denotes the look-back order and lookahead order of the memory block, respectively. $s_1$ is the stride factor of look-back filter and $s_2$ is the stride of lookahead filter.

Connectionist temporal classification (CTC) \cite{graves2006connectionist} is a loss function for sequence labeling problems that converts the sequence of labels with timing information into the shorter sequence of labels by removing timing and alignment information. The main idea is to introduce the additional CTC blank ($\textendash$) label during training, and then remove the blank labels and merging repeating labels to obtain the unique corresponding sequence during decoding. For a set of target labels, $\varOmega$, and its extended CTC target set is defined as $\bar{\varOmega}=\varOmega \cup \{\textendash\}$. Given an input sequence $\bf{x}$ and its corresponding output label sequence $\bf{y}$. The CTC path, $\pi$, is defined as a sequence over $\bar{\varOmega}$,  $\pi\in \bar{\varOmega}^T$, where $T$ is the length of the input sequence $\bf{x}$. The label sequence $\bf{y}$ can be represented by a set of  all possible CTC paths, $\Phi(\bf{y})$, that are mapped to $\bf{y}$ with a sequence to sequence mapping function $\mathcal{F}$, $\bf{y}=\mathcal{F}(\Phi(\bf{y}))$. Thereby, the log-likelihood of reference label sequence $\bf{y}$ given the input $\bf{x}$ can be calculated as an aggregation of the probabilities of all possible CTC paths:
\begin{equation}\label{eq.2}
p({\bf y}|{\bf x})=\sum_{\pi  \in \Phi(\bf{y})}p(\pi|\bf{x})
\end{equation}
Model training can then be carried out by minimizing the negative log-likelihood. Furthermore,  CTC trained acoustic model can be further optimized with sequence-level discriminative training criteria such as state-level minimum Bayes risk (sMBR) criterion \cite{sak2015acoustic,zhang2018deep}. 
\subsubsection{Acoustic modeling units}
In \cite{zhang2019deep}, it have investigated the performance of DFSMN-CTC-sMBR acoustic models with CI-IF, CD-IF, syllable and hybrid Character-Syllable as modeling units for Mandarin speech recognition. Experimental results suggest that the \emph{hybrid Character-Syllable} modeling units, which mixed the high frequency Chinese characters and syllables, is the best choice for Mandarin speech recognition. For \emph{hybrid Character-Syllable}, the low frequency characters are mapped into the syllables to deal with the OOV problem. In this work, instead of mapping the low frequency characters into syllables, we propose to map them into high frequency characters with the same pronunciation. As a result, we come up with a pure Chinese characters based modeling units without OOV for Mandarin speech recognition. Specifically, we keep the top 2000, 4000 and 6000 Chinese characters as acoustic modeling units, denoted as \emph{char-2k}, \emph{char-4k} and \emph{char-6k} respectively. The coverages is 95.58\%, 99.54\% and 99.86\%  in our text dataset, respectively.

\begin{table}[t]
	\centering
	\caption{Model architecture of the Transformers used in this work.}
	\begin{tabular}[t]{|c|cccccc|}
		\hline
		Transformer   & N & $d_{model}$ & $d_{ff}$ & h & $d_k$ & $d_v$ \\\hline
		small & 3 & 512         & 2048     & 4 & 512   & 512   \\\hline
		big   & 6 & 512         & 2048     & 8 & 512   & 512   \\\hline 
	\end{tabular}
	\label{tab:0}
\end{table}
\subsection{Decoder}
\subsubsection{Greedy search}
For greedy search \cite{graves2006connectionist} based decoding, the most likely symbol at each time step is chose as the output. The best CTC-path can be generated as followings:
\begin{equation}\label{eq.3}
\pi^*=\arg\max_{\pi} \prod_{t=1}^{N} P_{AM}(\pi_t|{\bf x})
\end{equation}
Furthermore, the CTC-path can mapped to the token sequence by using the mapping function $\mathcal{F}$.
\begin{equation}\label{eq.4}
{\bf y}=\mathcal{F}(\pi^*)
\end{equation}
For Chinese character units based CTC acoustic model, the token sequence is the final recognition results.  For syllable or other modeling units, it still need  another mapping function.
\subsubsection{WFST search}
The WFST-based decoding is proposed in \cite{miao2015eesen} that enables efficient incorporation of lexicons and language models into CTC decoding. The search graph is built by composing the language model WFST \emph{G}, lexicon WFST \emph{L} and token WFST \emph{T}. The token WFST T maps a sequence of frame-level CTC labels to a single lexicon unit. The overall oder of the FST operations is:
\begin{equation}\label{eq.5}
S=T\circ\min(\det(L \circ G))
\end{equation}
where $\circ$, $\det$ and $\min$ denote composition, determinization and minimization respectively. The search graph $S$ encodes the mapping from a sequence of CTC labels emitted on speech frames to the final transcription. The best decoding path can then be exported from the search graph $S$ using the beam search.
\subsubsection{N-best data expansion}
\label{sec:data_expansion}
 For both greedy search and WFST search based decoding, we usually take the best path as recognition result. In our work, we find the diversity of training data is important to the performance of spelling correction model. Thereby, we investigate to extend the diversity of training data using the N-best lists. For WFST-based decoding, we can easily get the top N paths from the decoding lattice. As to greedy search based decoding, we propose a threshold-based path retention method. At each time-step, in addition to retaining the token with the highest posterior probability ($p_1$), we judge whether to retain the second token (posterior probability, $p_2$) based on two thresholds ($upper\_th, lower\_th$). if $lower\_th<p_1<upper\_th$ and  $lower\_th<p_2$, we will keep both tokens. Otherwise, we only keep the top one token. Based on different thresholds, we will generate different CTC paths that can further mapped to token sequence with Eq.\ref{eq.4}.
 
 In our work, we also investigate to extend the training data using data augmentation with the text-only data. We try to add insertion, deletion and substitution errors to the original text based on a probability distribution. Unfortunately, this method doesn't work well. We suspect that constructed errors can't really simulate the types of error produced by the acoustic model.

\subsection{Speller}
For spelling correction model (\emph{speller}), we use the \emph{Transformer} with encode-decoder architecture, which is much similar to machine translation model in \cite{vaswani2017attention}. The preliminary recognition results generated by the front-end CTC-based acoustic models with different decoding and data expansion methods are used as input and the ground-truth transcriptions are used as output to train the speller. We use the OpenNMT toolkit \cite{opennmt} to train the Transformer based speller with default setting.
Specially, we have compared two Transformer architectures, denoted as \emph{small} and \emph{big}. The detailed configurations are as shown in Table \ref{tab:0}.
\newcommand{\pp}[1]{\raisebox{-1.2ex}[0pt][0pt]{\shortstack{#1}}}
\begin{table}[t]
	\centering
	\caption{CER (\%) of CTC based ASR systems with various acoustic modeling units and decoding methods (Greedy search Vs. WFST search).}
	\begin{tabular}[t]{|c|c|c|c|c|c|}
		\hline
		\pp{Exp} &\pp{Modeling units}  & \pp{Criterion} &  Greedy & WFST \\
		&                  &                &  Search  &   Search \\\hline
		1 & syllable &      CTC       &     -         &   5.55 \\\hline
		2 & char-2k  &      CTC       &    11.93      &   5.21 \\\hline
		3 & char-4k  &      CTC       &     \textbf{7.28}      &   5.20  \\\hline
		4 &          &      +SMBR     &     8.48      &   \textbf{4.42}  \\\hline
		5 & char-6k  &      CTC       &     7.33      &   5.25  \\\hline         
	\end{tabular}
	\label{tab:1}
\end{table}
\begin{table}[t]
	\centering
	\caption{Performance of greedy search CTC with speller. (1M: 1 million sentences.) }
	\begin{tabular}[t]{|c|c|c|c|c|}
		\hline
		Exp& Modeling units & Transformer &    Data     & CER\% \\\hline
		\pp{1} &\pp{syllable} & \pp{small}  &     1M   &  8.17 \\\cline{4-5}
		&             &             &     20M  &  6.40 \\\hline         
		2 &   char-2k    & small      &     20M  &  6.25 \\\hline
		3 &   char-4k    & small       &     20M  &  5.70 \\\hline
		4 &   char-4k    &	  big      &     20M  &  \bf{5.55} \\\hline
		5 &   char-6k    &    big      &     20M  &  5.65 \\\hline                
	\end{tabular}
	\label{tab:2}
\end{table}
\begin{table}[t]
	\centering
	\caption{Performance of speller trained with data generated by greedy search CTC and threshold-based data expansion.}
	\begin{footnotesize}
		\begin{tabular}[t]{|c|c|c|c|c|c|}
			\hline
			Training data & Steps/Pass     & Pass1   & Pass2 & Pass3 & Pass4 \\\hline
			D1        & 100000  & 5.55    & 5.18  & 5.02   & 4.89 \\\hline
			D1-3      & 200000   & 4.73    & 4.68  & 4.46   & 4.36 \\\hline
			D1-6      & 400000    & 4.62    & 4.38  & 4.28   & \textbf{4.21} \\\hline  
		\end{tabular}
		\label{tab:3}
	\end{footnotesize}
\end{table}
\begin{table}[t]
	\centering
	\caption{Performance of spellers trained with data generated by WFST search CTC and N-best data expansion.}
	\begin{footnotesize}
		\begin{tabular}[t]{|c|c|c|c|c|c|}
			\hline
			Training data  &  Steps/Pass & Pass1   & Pass2 & Pass3 & Pass4 \\\hline
			N-best(1)  & 100000 & 4.14   & 4.01  & 3.98   & 3.91 \\\hline
			N-best(5)  & 250000 & 3.79   & 3.69  & 3.60  & 3.54     \\\hline
			N-best(10) & 350000 & 3.72   & 3.61  &  3.50  & \textbf{3.41}  \\\hline
		\end{tabular}
	\end{footnotesize}
	\label{tab:5}
\end{table}
\section{Experiments}
\subsection{Experimental setup}
We conduct our experiments on a large Mandarin speech recognition task that consists of about 20,000 hours training data with about 20 million sentences.
A test set contains about 10 hours data is used to evaluated the performance of all models.  Acoustic feature used for all experiments are 80-dimensional log-mel filter-bank (FBK) energies computed on 25ms window with 10ms shift. We stack the consecutive frames within a context window of 5 (2+1+2) to produce the 400-dimensional features and then down-sample the inputs frame rate to 30ms. For WFST-based decoding, a pruned trigram language model trained with the text data is used. Evaluations are performed in term of character error rate (CER). For all experiments, we use the same DFSMN architecture as in \cite{zhang2018acoustic}.
CTC-based acoustic model is trained in a distributed manner using 16 GPUs and the Transformer based speller is trained using 2 GPUs. 
\subsection{CTC baseline system}
\label{sec:ctc_baseline}

The performance of various baseline CTC-based ASR systems with different acoustic modeling units and decoding methods are shown in Table \ref{tab:1}. CTC-based models decoded with WFST search perform much better than greedy search, which indicates the importance of linguistic information. Chinese character modeling units based CTC models with different numbers of characters (2k, 4k, 6k) can achieve similar performance when using WFST-based decoding, and all can outperform the syllable based CTC model. However, when decoded with greedy search, char-4k and char-6k based CTC models significantly outperform the char-2k based CTC model. This experimental result also indicates the importance of linguistic information. As to sMBR training, we find that it improves the performance when using WFST based decoding while hurt the performance when using greedy search based decoding. This is due to the mismatch between training and decoding, since sMBR based training uses the WFST-based decoding to generate the training lattices. 

\subsection{Greedy search CTC with speller}
\label{sec:greedy_ctc_speller}
First, we evaluate the performance of spellers trained with the output of baseline CTC models in Table \ref{tab:1} using greedy search based decoding. Comparison of the experimental results of \emph{exp2} and \emph{exp3} in Table \ref{tab:2} and Table \ref{tab:1} shows that good preliminary recognition result will lead to better final result. Moreover, the amount of training data is essential to the performance. As shown in Table \ref{tab:2}, increasing the training data from 1 million (1M) sentences to 20M sentences will result in more than 20\% relative improvement. Thereby, we also investigate how to extend the training data in this work. Since the \emph{big} Transformer based speller performs better than the \emph{small} one, we will use the \emph{big} one in our following experiments.

In Sec.\ref{sec:data_expansion}, we have introduced the threshold-based data expansion method for CTC-based acoustic model with greedy search. In this experiment, we use the baseline models in Table \ref{tab:2} with different thresholds to generate various training data, denoted as D(AM $upper\_th$ $lower\_th$). Specially, we use the CTC  model (\emph{exp3}) and CTC-sMBR model (\emph{exp4}) to generate six datasets, denoted as D1 (CTC 1.0 1.0), D2(CTC 0.5 0.1), D3(CTC 0.6 0.1), D4(CTC-sMBR 1.0 1.0), D5(CTC-sMBR 0.5 0.3), D6(CTC-sMBR 0.6 0.3). Different data configurations and experimental results are as shown in Table \ref{tab:3}. Inspired by the SGDR \cite{loshchilov2016sgdr}, we propose to train the speller with 4 passes and reset the learning rate after each pass. The training steps of each pass is determined by the amount of training data. For model inference, we use the results of CTC-based model with greedy search as input and generate the final recognition result by also using greedy search.  Results show that this optimization method will achieve a better convergence performance. More importantly, data expansion will significantly improve the performance of speller even  using the training data generated by CTC-sMBR model. As a result, we can achieve a character error rate (CER) of 4.21\%, which is 42.17\% relative improvement compared to the baseline greedy search decoded CTC model (\emph{exp3} in Table \ref{sec:ctc_baseline}). It can also outperform the CTC-sMBR model decoded with external language model using the WFST-based decoding. 
\begin{figure}[t]
	\centering
	\includegraphics[width=1.0\linewidth]{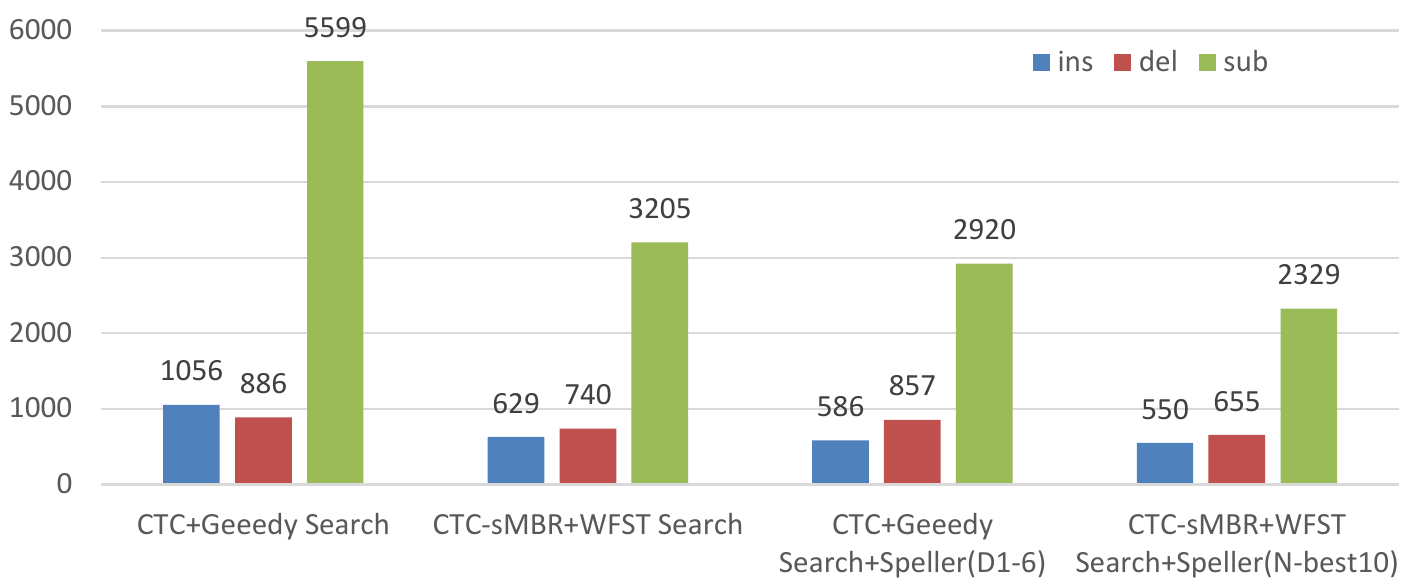}
	\caption{Error analysis of various systems.}
	\label{fig:error_type}
\end{figure}
\begin{figure}[t]
	\centering
	\includegraphics[width=1.0\linewidth]{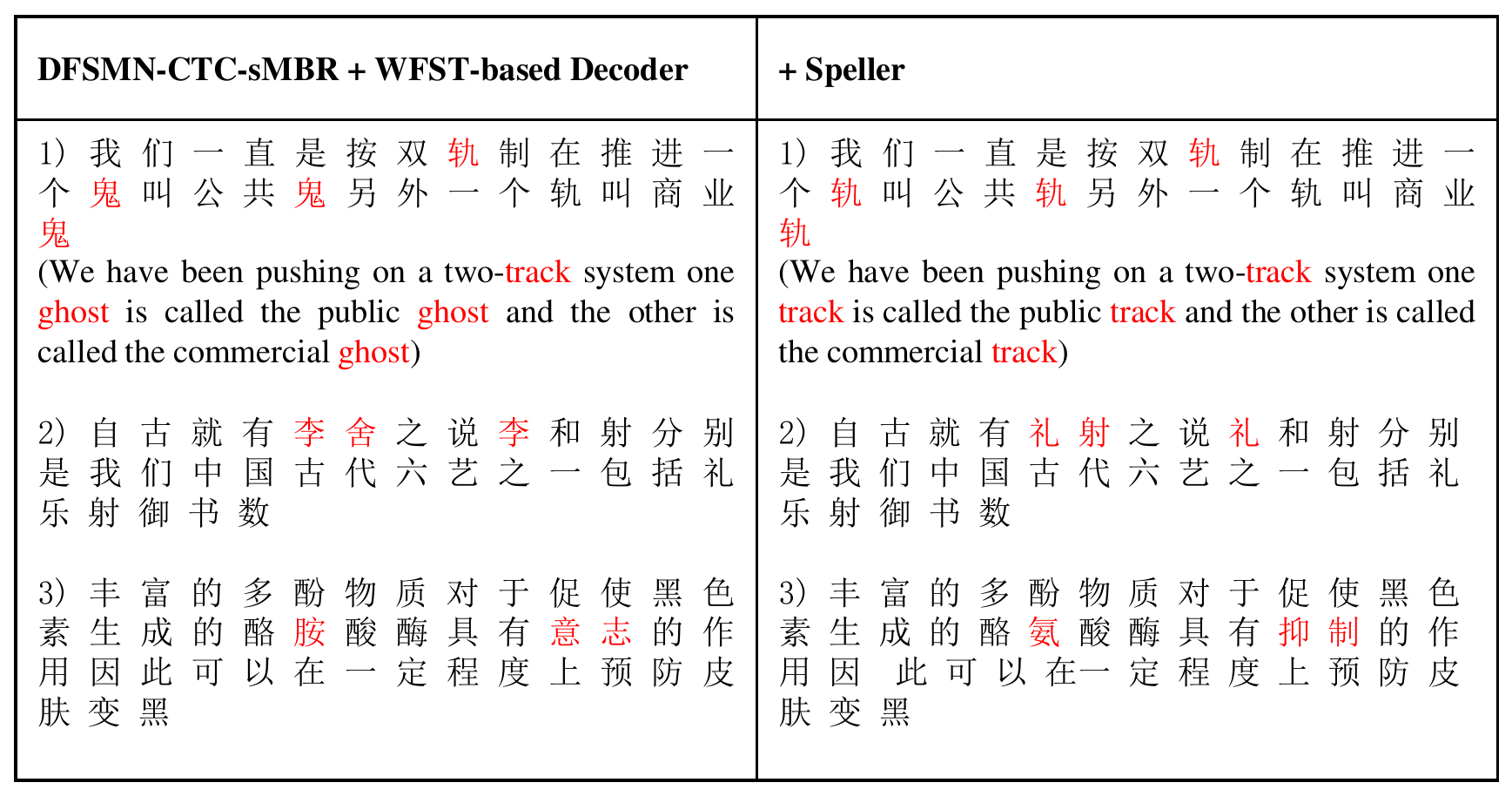}
	\caption{Examples of recognition result with and without speller.}
	\label{fig:example}
\end{figure}
\begin{figure*}[t]
	\centering
	\includegraphics[width=0.9\linewidth]{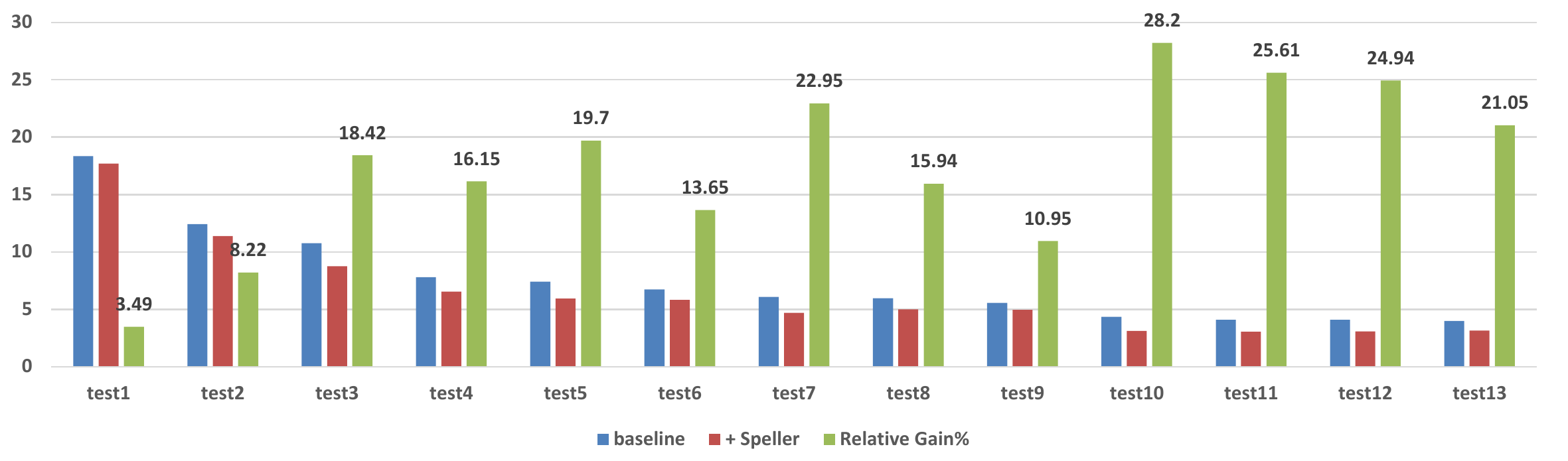}
	\caption{Performance of systems with and without speller in various test sets.}
	\label{fig:test_13}
\end{figure*}
\subsection{WFST search CTC with speller}
In this experiment, we use the DFSMN-CTC-sMBR listener (\emph{exp4} in Table \ref{sec:ctc_baseline}) with WFST-based decoder to generate the preliminary recognition results, which are then paired with the ground-truth transcriptions to train the speller. We also extend the training data with N-best lists in decoding lattice. Specially, we keep the top 1, 5 and 10 paths that result in three training sets, denoted as \emph{N-best(1)}, \emph{N-best(5)} and \emph{N-best(10)}. These three training sets contain of about 20 million (20M), 55M and 84M sentences respectively \footnote{There are less than 10 paths in the decoding lattice for many sentences in the training set.}. We list the experimental configurations and results in Table \ref{tab:5}.  During inference, the best paths from the WFST-based decoder are fed into the speller to generate the final recognition result with greedy search. The experimental conclusions are the same to greedy search CTC based experiments in Sec.\ref{sec:greedy_ctc_speller} that N-best based data expansion method and SGDR based multi-pass training can both significantly improve the performance. Finally, the speller trained with N-best(10) training set achieves a CER of 3.41\% while the performance of baseline DFSMN-CTC-sMBR system with WFST-based decoding is 4.42\%(in Table \ref{tab:1}).

\subsection{Analysis}
In order to understand the role of speller, as shown in Figure \ref{fig:error_type}, we plot the number of insertion, deletion and substitution errors in the recognition results of systems with and without speller. Results demonstrate that the speller can automatic correct many substitution errors made by the front-end listener no matter decoded with or without external language model. Figure \ref{fig:example} shows some representative examples in the test set. Results demonstrate that the speller is able to utilize the sentence-level linguistic information and learn knowledge base from the training set, which is helpful to distinguish homophone in Mandarin. Moreover, we also make an extensive evaluation of the speller. Figure \ref{fig:test_13} shows the performance of baseline DFSMN-CTC-sMBR system with and without speller in 13 test sets. We sort the test sets with the CER(\%) of baseline system from high to low. Results show that usually the better the baseline, the more performance gains you can achieve by using speller. This is because speller needs to utilize the context information. If there exists many errors in the original recognition results, it will  
increase the difficulty of error correction.

\section{Conclusions}
In this work, we propose a transformer based spelling correction model with encode-decoder architecture to automatically correct errors made by CTC-based speech recognition system. Experimental results show that the speller is able to utilize the sentence-level linguistic information, which will help to significantly reduce the substitution errors in the recognition results. Moreover, we propose to extend the diversity of training data using the N-best based data expansion method that results in more than 10\% relative improvement. Results in a 20,000 hours Mandarin speech recognition task show that the proposed spelling correction model can achieve a CER of 3.41\%, which results in 22.9\% and 53.2\% relative improvement compared to the baseline CTC-based systems decoded with and without language model respectively. As to future work, we will investigate to utilize the text-only data that will significantly extend the diversity of training data.

\bibliographystyle{IEEEtran}

\bibliography{mybib}
	
\end{document}